\documentclass[particles,article,submit,moreauthors,pdftex]{Definitions/mdpi} 

\firstpage{1} 
\pubvolume{xx}
\issuenum{1}
\articlenumber{5}
\pubyear{2019}
\copyrightyear{2020}
\history{Received: 28 February 2020; Accepted: 1 April 2020; Published: date}

\pdfoutput=1

\usepackage{amssymb}
\usepackage{slashed}
\usepackage{mathtools}
\usepackage{feynmf}
\usepackage{mathrsfs}
\usepackage{simpler-wick}

\usepackage[export]{adjustbox}
\usepackage[colorinlistoftodos]{todonotes}
\DeclareMathOperator{\Lagr}{\mathscr{L}}
\usepackage{subfigure}
\usepackage{indentfirst} 

\graphicspath{{Pics/}}

\Title{Cosmic Gamma Ray Constraints on the Indirect Effects of Dark Matter
}

\Author{Konstantin M. Belotsky, Airat Kh. Kamaletdinov, Ekaterina S. Shlepkina * and Maxim~L.~Solovyov}

\AuthorNames{K.M. Belotsky,  A.Kh. Kamaletdinov, E.S. Shlepkina  and  M.L. Solovyov}

\address[1]{%
Moscow Engineering Physics Institute, National Research Nuclear University MEPhI, Kashirskoe Shosse 31, 115409 Moscow, Russia; k-belotsky@yandex.ru  (K.M.B.); kamaletdinov.a.h@yandex.ru (A.K.K.); max07s@mail.ru~(M.L.S.)\\
}

\corres{ shlepkinaes@gmail.com; Tel.: +7-980-512-2503}

\abstract{The observed anomalous excess of high-energy cosmic ray (CR) positrons is widely discussed as possible indirect evidence for dark matter (DM). However, any source of cosmic positrons is inevitably the source of gamma radiation.
The least model dependent test of CR anomalies interpretation via DM particles decays (or annihilation) is connected with gamma-ray background due to gamma overproduction in such processes.
In this work, we impose an observational constraint on gamma ray production from DM.
Then, we study the possible suppression of gamma yield in the DM decays into identical final fermions. Such DM particles arise in the multi-component dark atom model.
The influence of the interaction vertices on the gamma suppression was also considered.
No~essential gamma suppression effects are found.
However, some minor ones are revealed.
}

\begin{document}


\section{Introduction}

{The physical nature of DM still is unknown and it s obviously connected with physics beyond the Standard Model (SM)} 
{ Many DM search attempts, both direct in underground experiments and accelerator ones and indirect in cosmic ray (CR) experiments  
are undertaken}.
{We continue our investigations in this field} \cite{Belotsky:2016tja,1742-6596-675-1-012023,1742-6596-675-1-012026,Alekseev2017An,Belotsky:2017wgi,Alekseev2017R,Belotsky:2018vyt,2019PDU....2600333B,2019IJMPD..2841011B,ICPPA-2019}.

This {note} is {devoted to the} explanation of {cosmic positron excess in the AMS-02 data} \cite{PhysRevLett.113.121101} with the help of DM annihilation or decay in Galaxy.{ As it was shown previously, the~main problem lies in the predicted contradiction with the data in gamma-ray background obtained by Fermi-LAT~\cite{Ackermann:2014usa}. Many things have already been done in this field. This article considers several new mechanisms of gamma-ray suppression and estimates the necessary level of such suppression.}

{The article structure is the following. First we estimate maximal possible  photon yield in DM particle decay/annihilation from observational data (Section \ref{Max}). Gamma-ray constraints have not been presented in such terms earlier. Then we assess effect of final state radiation (FSR) suppression due to the so-called Single-Photon theorem (Section \ref{Single}). This is first studied in this article. We also  consider the refined Lagrangian impact on possible photon yield suppression. For~this purpose we compare at analytic level the scalar and vector cases of DM particle coupling, and~consider the coupling with derivative, which have never been done previously (Section \ref{A}). 
{Section~\ref{Met} clarifies in greater detail what has been done in this work as compared to the previous ones, and how it was accomplished.} 
}
 
 \label{IN}

\section{Methods}
\label{Met}
{Here, we continue our investigation of possibility to suppress the photon yield in the process of DM decay or annihilation in the framework of the simplest models of DM interaction with ordinary matter. We focus on the interactions which lead to the decays of DM particles into electrons and positrons. Such interactions can be described by Lagrangian \eqref{eq:SimplestV} for scalar DM particle decays and \eqref{eq:SimplestVect} for vector DM particle case which are given below. Other types of decays can also be considered, however, we are interested in the simplest (supposedly minimal) case of photon production, which is enough to study the possibility of gamma suppression. Moreover, more complicated decays cannot be analytically calculated to obtain the explicit dependence of the photon yield on the model parameterization. 
The~goal was to check the possibility of achieving the photon yield suppression due to parameter variation explicitly. Methods used here are partially taken from our previous works. 
}

{
In this article, we give analytic expressions of the corresponding decay widths and compare them with each other. Here we are looking for a difference in the energy distributions of the considering decay widths in order to determine in which case (scalar or vector) the high-energy photon yield will be lower. Thus, we are talking about suppression due to final state photon energy instead of suppression due to Lagrangian parameterization, considered in our previous works.}

{We also continue our search of the photon yield suppression due to Lagrangian parameterization and therefore introduce a new type of vertices containing special derivative terms. We show that such vertices lead to parameter-dependent energy distributions of photon production
(unlike the vertices considered in previous works). This dependence allows us to minimize the photon yield by varying these model parameters.
}

{
In addition, we propose the new approach to possible suppression of FSR by using the identical final fermions in decay of DM particles. The~fact is that, in the classical case, the system of identical charged particles do not radiate at all (dipole radiation is zero). In~the quantum case, there is the so-called single-photon theorem which tells about (at least) partial suppression of photon radiation under some~conditions. 

}

{For estimation of the maximal possible photon yield from viewpoint of the least tension with observation data on cosmic gamma-rays, we make use of chi-square and its minimization. According to our previous results, the~annihilation of DM particles with mass $M$ and decay of DM particles with mass $2M$ lead to almost the same results. Therefore, for~this task we consider the annihilation case only. More details are given in the Section~\ref{Max}}.

The tools which we use in our calculations were basically described in our previous articles, for~example,~\cite{,Belotsky:2016tja, 2019IJMPD..2841011B}. For~the calculation of CR propagation we use the GALPROP code~\cite{GALPROP_eng}. We also use HEP MC-generators and program software like CalcHEP~\cite{belyaev2013calchep}, LanHEP~\cite{semenov1998lanhep} for calculation of the spectrum of the products of DM particle annihilation or decay. For~analysis of final results we used standard mathematical programs like Wolfram Mathemathica and its specialized packages like FeynCalc and FeynRules~\cite{christensen2009feynrules}.

\section{Results}
\label{Res}

\subsection{Maximal possible gamma-ray radiation from annihilating DM model explaining the positron excess in CR}
\label{Max}
In our works~\cite{1742-6596-675-1-012023,1742-6596-675-1-012026,Belotsky:2016tja,Belotsky:2017wgi, 2019IJMPD..2841011B} 
we have shown that the DM models explaining the positron anomaly in most cases have major problems with gamma-ray constraints, more specifically, the~one set by IGRB data obtained by Fermi-LAT experiment~\cite{Ackermann:2014usa}. And~therefore, to~make these models plausible, one needs to somehow suppress the gamma-ray emission. In~one of our last works~\cite{2019PDU....2600333B} we have tried some possible techniques, but~they proved to be ineffective. So in this work, we decided to make a simple estimation for necessary gamma-ray~suppression. 

Here we consider  the AMS-02~\cite{PhysRevLett.113.121101} data on the positron fraction and the basic leptophilic DM model with DM particle being able to annihilate into lepton-antilepton pairs. The~averaged over speed cross-section, as~well as branching ratios of the leptonic channels ($e^+e^-,\;\mu^+\mu^-,\;\tau^+\tau^-$) are parameters of the model and are obtained by minimizing the chi-square. To~obtain our estimation, however, we modify our usual expression for the chi-square:
\begin{gather}
    \chi^2= \left[\, \sum_{\substack{\rm AMS-02}}\frac{\left(\Phi_{e^+}^{th}-\Phi_{e^+}^{obs}\right)^2}{\sigma_e^2}+\sum_{\substack{\rm Fermi}} \frac{\left(K\Phi_{\gamma}^{th}-\Phi_{\gamma}^{obs}\right)^2}{\sigma_{\gamma}^2}\,\Theta\left(K\Phi_{\gamma}^{th}-\Phi_{\gamma}^{obs}\right) \right].
    \label{chi2}
\end{gather}
Here  $\Phi_{i}$ are the predicted (\textit{th}) and measured (\textit{obs}) fluxes for {$i = e^+,\gamma$} denoting the positron fraction or gamma datapoints respectively, $\sigma_{i}$ denotes the corresponding experimental errors, $\Theta$ is the Heaviside theta-function to ensure we do not go over the experimental limits. We should emphasize that the coefficient $K$, representing gamma suppression, has been introduced for the first time in this work. 
The~first sum in Equation~\eqref{chi2} goes over the AMS-02 data points and the second sum goes over the Fermi-LAT datapoints. AMS points are taken in the range $30\div 500$ GeV,{ and Fermi ones \mbox{$30\div600$ GeV}}. 

However, it is clear that the best fit would be achieved with no gamma at all, i.e.,~with $K=0$. Therefore, to~obtain an estimation we instead search for the solution of the equation:
\begin{equation}
    \chi^2 = \chi^2_{p, N},
    \label{chi_quant}
\end{equation}
where $\chi^2_{p, N}$ is the quantile for chi-square distribution with p-value $p=0.01$ and degrees of freedom $N$, which includes all used AMS-02 datapoints and those Fermi datapoints, where we have excess over the experimental~data. 

The analysis goes as follows. We start with minimizing the first sum in expression \eqref{chi2} to obtain the best-fit for positron anomaly and parameters (except K) that allow it. Then we proceed to solve Equation \eqref{chi_quant} with obtained parameters being~fixed.

\begin{figure}[h]
    \centering
    \includegraphics[width=0.9\textwidth]{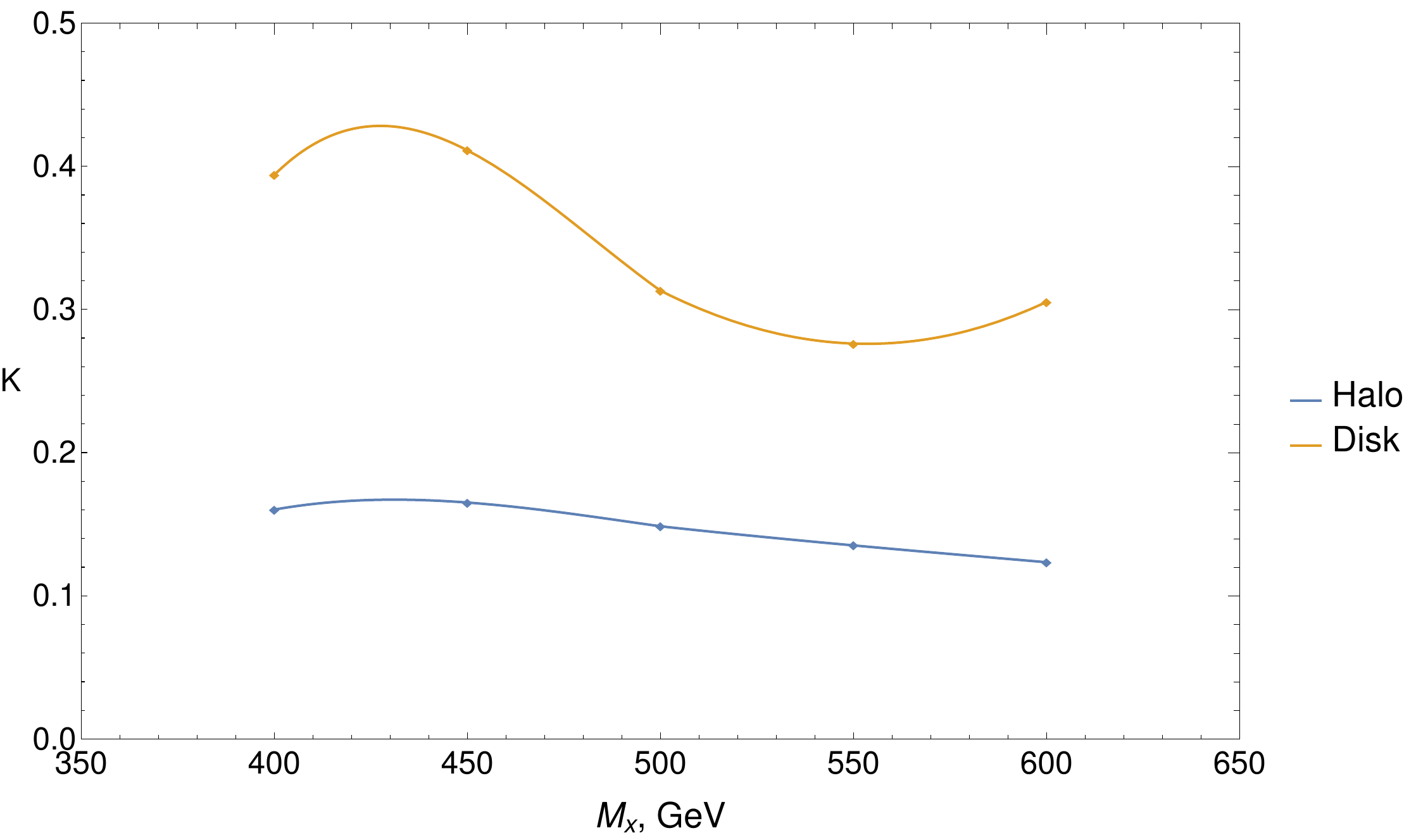}
    \caption{The suppression coefficient K in dependence of mass of initial DM particle}
    \label{gk}
\end{figure}

We have conducted the analysis for the model of an annihilating halo of DM (with the NFW density profile~\cite{Navarro:1996gj}) and for the dark disk we {work on} (with Read's density profile~\cite{2008MNRAS.389.1041R}). The~results are presented in the Figure~\ref{gk}.
    
It must be noted, though, that the dark disk model was proposed to lessen the contradiction with IGRB data in simultaneous {(psoitrons + gamma)} fit  {at the cost of} slightly worse positron fit. 
{And in the case we consider now, as~we suppress gamma artificially, this slightly worse positron fit becomes a problem}. 
Read's density profile causes a lack of low and mid-energy positrons leading to high values of chi-square even without gamma at all. At~higher masses of initial particle it results in absence of solutions for Equation \eqref{chi_quant}, and~in that case the closest obtainable value was used. Therefore, the~K values obtained for the disk case should be treated with caution. However, it seems to be possible to obtain proper estimations with additional fit of disk thickness for every mass~particularly.

\subsection{On the suppression due to single-photon theorem}

{Earlier we have noted that the mode with identical final fermions like $X\rightarrow e^+e^+(\gamma)$ has an advantage with respect to 
$e^+e^-$ modes from viewpoint of FSR suppression for a simple reason---we have two positions in each reaction in contrast with the case of $X \rightarrow e^+ e^-(\gamma)$. Thus, one can obtain the same number of positrons and lessen the gamma production by half. Such models were considered in~\cite{Khlopov2006,Belotsky:2014haa,khlopov2006dark,belotsky2014dark}, including the context of cosmic positron anomaly solution.
Here we are going to consider a more refined effect of an identity of final fermions which can additionally contribute to this suppression. We have mentioned above that full gamma (dipole)-radiation suppression takes place in the classical case. In~quantum case, similar effect seems to take place in some degree as Single-Photon theorem tells, which we consider in this work.}

\label{Single}
\subsubsection{Considering models}

We study models of dark matter, consisting of hypothetical long-lived scalar or vector particles $X$, with~masses about 1--3 TeV. As~mentioned above, we consider leptonic decay modes. 
Mostly, two dark matter particle models were considered:

\begin{enumerate}[leftmargin=*,labelsep=4.9mm]
\item The simplest model of dark matter particle decay into two oppositely charged leptons ( $X\rightarrow~e^+~e^-$ and  $X\rightarrow~e^+~e^-\gamma$):
\begin{equation}
    \Lagr = X\overline{\psi} (a+b\gamma^5)\psi +  \overline{\psi}\gamma^{\mu}A_{\mu}\psi,
    \label{eq:SimplestV}
\end{equation}
\begin{equation}
    \Lagr = X_{\mu}\overline{\psi} \gamma^{\mu} (a+b\gamma^5)\psi +  \overline{\psi}\gamma^{\mu}A_{\mu}\psi.
    \label{eq:SimplestVect}
\end{equation}

\item The model of decay of a dark matter particle into two identical positrons ( $X\rightarrow~e^+~e^+$ and  $X\rightarrow~e^+~e^+\gamma$). Such models were proposed and studied in~\cite{Khlopov2006,Belotsky:2014haa,khlopov2006dark,belotsky2014dark}:
\begin{equation}
    \Lagr = X\overline{\psi^C} (a+b\gamma^5)\psi + X^*\overline{\psi} (a-b\gamma^5)\psi^C - \overline{\psi}\gamma^{\mu}A_{\mu}\psi
\end{equation}
where $a$ and $b$ are the arbitrary model parameters{, and~everywhere $\psi$ and $\psi^C$ are the fermion wave function and its charge conjugated one respectively.}
\end{enumerate}

\subsubsection{Contribution to the suppression effect by the identity of particles in the final state}

To understand whether the contribution to the photon suppression effect is the result of the identity of the final state particles, one can obtain 
the ratio of branching ratios of three-body decay of dark matter particles  with identical fermions in the final state over   branching ratio of the simplest electron-positron~mode:
\begin{equation}
 \label{eq:BR}
\frac{Br (X \rightarrow e^+ e^+ \gamma)}{Br(X \rightarrow e^+ e^-\gamma)}  \rightarrow \rm min. \newline
\end{equation}

{There can be some} limited range of physical parameters in which suppression due to this effect could be observed. This phenomenon is described by the so-called Single-Photon theorem (or radiation zeros), which is considered in~\cite{brown1995understanding,doncheski1998radiation,stirling2000radiation,heyssler1998radiation}.
Thus, one can trace the dependence effect of suppression of the photon yield on model parameters. If~the contribution of positron identity in the final state is made, then dips will appear on such dependencies. {It is expected that there can be dips in dependencies of Equation~\eqref{eq:BR} from parameters as well as of differential probability of the process itself from kinematic~parameters.}

An analysis of the branching ratio was carried out depending on {the following} parameters:
DM particle mass, the energy of emitted photon and angle between photon and lepton. The~probabilities of the mentioned processes were considered in differential form, in dependence on the energy and angle. 
{For calculation, CalcHEP code was used. Cut on photon energy was imposed accepting only $\omega>1$~GeV (to circumvent infrared divergence problem).}

It was found that it is hard to pick out the region of values of DM particle mass or emitted photon energy, where there would be suppression of the photon due to the identical lepton in the final~state. 

\begin{figure}[h]
    \centering
    \includegraphics[width=0.55\textwidth]{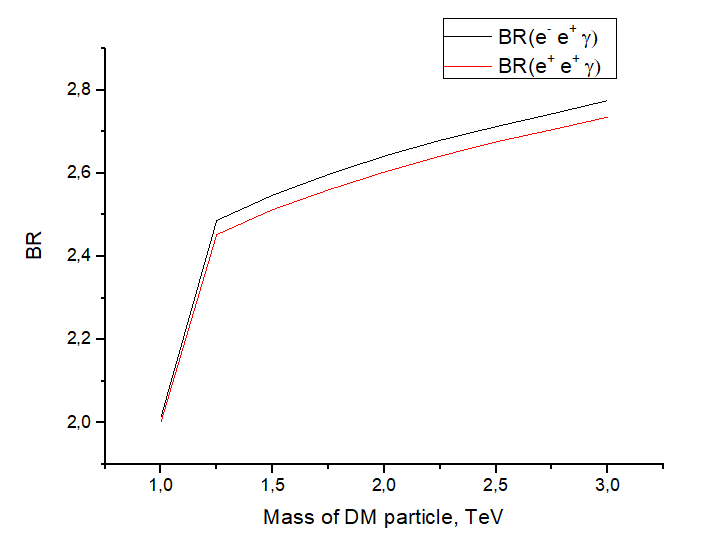}
    \caption{Dependence of $Br(e^+ e^+ \gamma)$ and $Br(e^+ e^-\gamma)$ on the mass of the initial particle.}
    \label{BRM}
\end{figure}

This is shown in Figure~\ref{BRM}{ for dependence on DM particle mass}. As~one can see the difference in branching ratios of these processes is not significant and their ratio will approach one ``1``. The~situation is the same as the case of the emitted photon energy of~. 

\begin{figure}[H]
    \centering
    \includegraphics[width=0.55\textwidth]{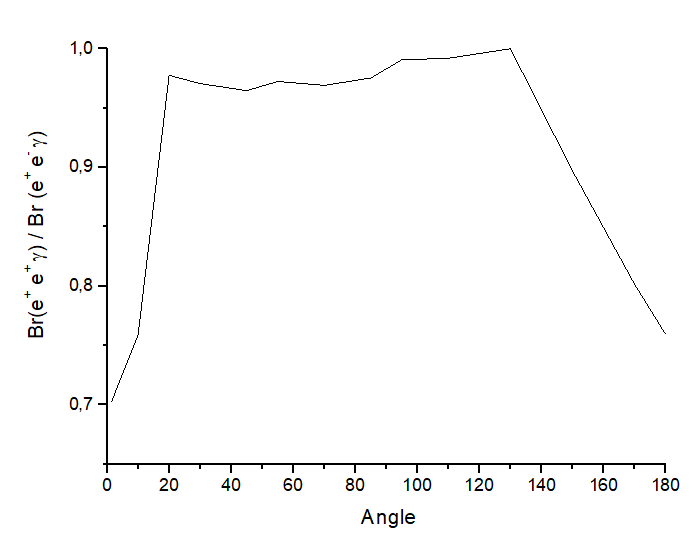}
    \caption{Dependence of $Br(e^+ e^+ \gamma)/Br(e^+ e^-\gamma)$ on the scattering angle between particles.}
    \label{BRA}
\end{figure}

In~\cite{brown1995understanding} examples were given in which ``radiation zeros'' appeared. It was experimentally registered by dips in angular distributions.
Further, we've studied the dependence of the branching ratio on the angle between the photon and the particle emitted it. This dependence is shown in the Figure~\ref{BRA}.
{The fractured behavior of the curve in Figure~\ref{BRA} is due to limited statistics and an internal error in the software we have used.}

From this dependence, one can conclude that there is a small range of the angles, within~which there is some suppression of the photon yield (about 30\%). It means that the effect of the final lepton identity {does not seem to play a role in our CR task.}

\subsection{On the suppression due to DM/SM interaction Lagrangian}
\label{A}

As we mentioned above, one can try to suppress photon yield due to the interaction Lagrangian of DM particles. Earlier we have shown~\cite{2019IJMPD..2841011B} that the parametrization of the simplest interaction vertices like \eqref{eq:SimplestV} and \eqref{eq:SimplestVect} does not suppress the photon yield during decays of vector and scalar DM particles into~$e^+ e^-$.

{In this part we want to show the influence of the spin of the decaying DM particle on the final state high-energy photon yield. For~this purpose, analytical expressions of the DM particle differential decay widths depending on the FSR photon energy $\omega$ were obtained.}

\begin{figure}[h!]
    \centering
    \includegraphics[width=0.85\linewidth]{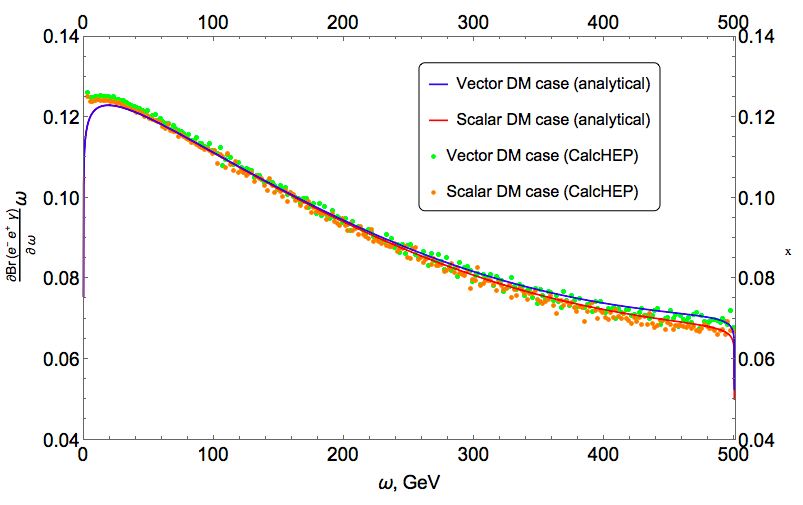}
    \caption{The photon energy distributions of $\cfrac{\partial Br(e^- e^+ \gamma)}{\partial \omega}\cdot\omega$ for the case of scalar DM particle (red) and vector DM particle (blue) decay with corresponding distributions obtained in CalcHEP. For $m = 1000$ GeV.}
    \label{fig:ScalarVectorDistrs}
\end{figure}

{Here, we present the result of comparing such analytical expressions for scalar DM particles \eqref{eq:ScalarDistr} and for vector ones \eqref{eq:VectorDistr} and compare them with the corresponding distributions obtained in CalcHEP (see Figure~\ref{fig:ScalarVectorDistrs}).}

For scalar DM particle case $\Lagr = X \overline{\psi}\psi$, $\Lagr = X \overline{\psi}\gamma^5\psi$,
we have calculated:
\begin{equation}
    \frac{\partial Br(e^- e^+ \gamma)}{\partial \omega} = -\frac{e^2 (m^2 - 2 m \omega + 2\omega^2)\ln(|\frac{m-2E_{\rm{e}}}{m-2(E_{\rm{e}}+\omega)}|)}{4 \pi^2 m^2 \omega}\Biggr|_{E_{\rm{e}}^-}^{E_{\rm{e}}^+}.
    \label{eq:ScalarDistr}
\end{equation}

For vector DM particle case $\Lagr = X_{\mu} \overline{\psi}\gamma^{\mu}\psi$, $\Lagr = X_{\mu} \overline{\psi}\gamma^{\mu}\gamma^5\psi$ we have:
\begin{equation}
    \frac{\partial Br(e^- e^+ \gamma)}{\partial \omega} = -e^2\frac{(m^2 - 2 m \omega + 2\omega^2)\ln(|\frac{m-2E_{\rm{e}}}{m-2(E_{\rm{e}}+\omega)}|) - 4 E_{\rm{e}} \omega}{4 \pi^2 m^2 \omega}\Biggr|_{E_{\rm{e}}^-}^{E_{\rm{e}}^+},
    \label{eq:VectorDistr}
\end{equation}
where $E_{\rm{e}}^+$ and $E_{\rm{e}}^-$ is the upper and lower kinematic limits of the electron in $X \rightarrow e^+ e^- \gamma$ decay and $m$ is the mass of the $X$ particle. Corresponding distributions are shown in Figure~\ref{fig:ScalarVectorDistrs}.

In these considered cases, the~following distributions are turned out to be independent on the parametrization of Lagrangian \eqref{eq:SimplestV} and \eqref{eq:SimplestVect}.
However, it can be seen that there is a {slight} suppression of the photon yield in the decays of scalar dark matter particles in comparison with the vector ones {at high energies}.
Unfortunately, such suppression is not sufficient to resolve the contradiction between the gamma-ray observation data and corresponding theoretical~predictions.

{In order to obtain the dependence of photon production on the model parameters $a$, $b$ we introduce the derivative factor in the Lagrangian }
\begin{equation}
    \Lagr = \overline{\psi} \gamma^{\mu} \Big(a + \frac{i b (\gamma^{\nu} \partial_{\nu})}{m}\Big) X_{\mu} \psi
    \label{eq:NewLagr} .
\end{equation} 

{This derivative factor leads to the difference between parameter-dependence of two-body and three-body decays (which could not be achieved in previous works).}

{An analytic distribution of the DM particle differential decay width for this model as a function of FSR photon energy is given below.}

\begin{equation}
    \frac{\partial Br(e^- e^+ \gamma)}{\partial \omega} = e^2\frac{m (2 a^2 + b^2) (m^2 - 2 m \omega + 2\omega^2)\ln(|\frac{m-2E_{\rm{e}}}{m-2(E_{\rm{e}}+\omega)}|) - 8 E_{\rm{e}} \omega (a^2 m + 2 b^2 \omega)}{-4 \pi^2 m^3 \omega (2 a^2 + b^2)}\Biggr|_{E_{\rm{e}}^-}^{E_{\rm{e}}^+}.
    \label{eq:DependedDistr}
\end{equation}

{The dependence of the photon yield on the model parameters during decays of DM particles allows one to search for final state gamma suppression by varying such parameters.}

{However, this also does not seem to be sufficient for the photon suppression in explanation of cosmic positron anomaly with DM.}

\section{Conclusions}

In this work, we estimated the possible branching ratio of photon yield in DM particles annihilation or decay in order to not contradict the data on cosmic gamma-ray background. {We~show again, the~halo DM models are highly unfavorable, while dark disk model is slightly more promising, but~it needs the more detailed analysis.}

We also discussed the possibility of suppression of photon yield due to the so-called Single-Photon theorem for the case when we have identical charged fermions in final states {(two positrons)}. There~is the model which allows this. We obtained the respective plot for these values in dependence on DM particle mass and angles. It was shown that the effect seems to be small. It was anticipated that the effect of suppression, in this case, should follow the classical case (dipole radiation {vanishes} {for system of} two equally charged particles). Nonetheless, it was found to be weak in the quantum case. {But we draw attention to this option of photon suppression which, maybe, should be investigated more.}

The extra possible Lagrangian effect was also considered. We considered the cases of scalar and vector DM particles and compared them with each other and with the results of the HEP MC-generator. Also~took the Lagrangian with derivative term. No essential effect of photon suppression was found~yet.

\vspace{6pt}

\textit{Acknowledgments}

We would like to thank R.I.Budaev, M.N.Laletin for background created by them to this work, V.A.Lensky for theoretical support, M.Yu.Khlopov, S.G.Rubin for interest to the work with useful discussion and A.A.Kirillov for technical assistance.

The work was supported by the Ministry of Education and Science of the Russian Federation, MEPhI Academic Excellence Project (contract  02.a03.21.0005, 27.08.2013). The~work of K.B.\ is also funded by the Ministry of Education and Science of the Russia, Project  3.6760.2017/BY.
The work of E.Sh.\ was also supported by MDPI ``Particles'' grant ``for student’s scientific~debut''.

\reftitle{References}
\externalbibliography{yes}

\bibliography{Bibliography}

\end{document}